\documentclass[aps,a4paper,12pt, twocolumns]{revtex4}
\usepackage{float}
\usepackage{fancyhdr}
\usepackage{color}
\usepackage{graphicx}
\usepackage{epsfig} 
\usepackage{amssymb}
\usepackage{amsmath,amsfonts}
\usepackage{booktabs}
\usepackage{hyperref}
\usepackage{pgfplotstable}
\usepackage[footnotesize,singlelinecheck=off,justification=raggedright]{caption}
\def\bra#1{\mathinner{\langle{#1}|}}
\def\ket#1{\mathinner{|{#1}\rangle}}

\def\Bra#1{\left<1>}
\DeclareMathAlphabet{\mathbbmsl}{U}{bbm}{m}{sl}
{\catcode`\|=\active\gdef\Braket#1{\left<\mathcode`\|"8000\let|\bravert {#1}\right>}}
\def\bravert{\egroup\,\vrule\,\bgroup}

\def\Tr{\mathop{\mbox{\normalfont Tr}}\nolimits}

\begin{document}

\title{Entanglement in fermionic chains with finite range 
coupling and broken symmetries}

\author{Filiberto Ares\footnote{Corresponding author.}}
\email{ares@unizar.es}
\affiliation{Departamento de F\'{\i}sica Te\'orica, Universidad de Zaragoza,
50009 Zaragoza, Spain}
\author{Jos\'e G. Esteve}
 \email{esteve@unizar.es}
 \affiliation{Departamento de F\'{\i}sica Te\'orica, Universidad de Zaragoza,
50009 Zaragoza, Spain}
\affiliation{Instituto de Biocomputaci\'on y F\'{\i}sica de Sistemas
Complejos (BIFI), 50009 Zaragoza, Spain}
  \author{Fernando Falceto}
\email{falceto@unizar.es}
 \affiliation{Departamento de F\'{\i}sica Te\'orica, Universidad de Zaragoza,
50009 Zaragoza, Spain}
\affiliation{Instituto de Biocomputaci\'on y F\'{\i}sica de Sistemas
Complejos (BIFI), 50009 Zaragoza, Spain}
  \author{Amilcar R. de Queiroz}
\email{amilcarq@gmail.com}
 \affiliation{Departamento de F\'{\i}sica Te\'orica, Universidad de Zaragoza,
50009 Zaragoza, Spain}
 \affiliation{Instituto de Fisica, Universidade de Brasilia, Caixa Postal 04455, 70919-970, Bras\'{\i}lia, DF, Brazil}


\begin{abstract} 
 We obtain a formula for the determinant of a block Toeplitz matrix 
 associated with a quadratic fermionic chain with complex coupling. 
 Such couplings break reflection symmetry and/or charge conjugation symmetry. 
 We then apply this formula to compute the R\'enyi entropy of a partial 
 observation to a subsystem consisting of $X$ contiguous sites in the limit of large $X$. 
 The present work generalizes similar results due to Its, Jin, Korepin and Its, Mezzadri, Mo. 
 A striking new feature of our formula for the entanglement entropy 
 is the appearance of a term scaling with the logarithm of the size of $X$. 
 This logarithmic behaviour originates from certain discontinuities 
 in the symbol of the block Toeplitz matrix. Equipped with this formula we 
 analyse the entanglement entropy of a Dzyaloshinski-Moriya spin chain and
 a Kitaev fermionic chain with long range pairing. 
\end{abstract}

\maketitle

\section{Introduction}

The ability of controlling parts or subsystems of a 
quantum system is the main resource of a future quantum computer 
with its processing superiority over nowadays computers. 
Such ability presupposes the control of local operations on parts of 
the system and its effects due to quantum correlations on other parts of 
the system. Such quantum correlations among distinct parts of the 
system are encoded by the keyword entanglement -- the characteristic trait of 
quantum mechanics, in the words of Schr\"odinger. 
It is therefore of great interest to be able to quantify or measure 
how ``entangled'' the parts of the system are. 
This is the reason of the great effort being carried in 
the past decades to elucidate the notion of quantum entanglement.

Entanglement has been studied in diverse systems, 
the simplest and most studied being those associated with quadratic 
spinless fermionic chains. These are related via a (non-local) 
Jordan-Wigner transformation to simple spin systems such as Ising, 
XX and XY with an external magnetic coupling. 
It is fair to say that nowadays we have a good control of 
almost all aspects of quantum correlations in such systems. 
Such achievement were possible due to the sophisticate resolution 
of diverse mathematical problems, most of them associated with the 
computation of the determinant of correlation matrices -- usually of the 
block Toeplitz type due to translational symmetry. 
The sophistication may be appreciated by noting that such computations 
are instances of the Riemann-Hilbert problem (RHP). 
This fact brings in a geometrical flavour to the study of 
entanglement entropy for such systems. 
It is a question of the analysis of the topology of a certain Riemann 
surface associated to the system. 

In the framework of this successful ideology, we can find two
classes of block Toeplitz matrices \cite{Deift}. 
On one hand, those which can be reduced to a
pure Toeplitz one, with scalar symbol, where the Fisher-Hartwig
theorem \cite{Basor} is the key to obtain an asymptotic expansion of our determinant. That is the case 
of spinless fermionic chains described by a Hamiltonian which preserves the fermionic number.
In this way, Jin and Korepin \cite{Jin} obtained a formula for the von Neumann entropy
associated to a set of contiguous sites in the XX spin chain. Using the same 
philosophy, the R\'enyi entropy of an interval for every stationary state
of a general quadratic spinless fermionic Hamiltonian which preserves the number of excitations
is obtained in \cite{Ares1}. 
On the other hand, we have those block Toeplitz matrices which cannot be reduced
to a scalar one; as usually happens when the above number symmetry is broken. 
In general, we cannot bypass the RHP. Confronting it, Its, Jin and Korepin \cite{Its}
obtained a formula for the von Neumann entanglement
entropy associated to a set of contiguous sites in the XY spin chain. 
An expression for the R\'enyi entropy 
was then written down by Franchini, Its and Korepin \cite{Franchini2}. 
Later the von Neumann entropy for the case of quadratic 
spinless fermionic chains with finite range coupling was provided by Its, Mezzadri and Mo \cite{Its2}. 
In a different spirit, using a duality transformation, Peschel \cite{Peschel2} obtained the same kind of formula
for the XY spin chain.

Later on, K\'adar and Zimbor\'as \cite{Kadar} started a systematic analysis of the entanglement 
entropy for self-dual models with broken reflection symmetry. 
It is interesting to note that in this self-dual case the block Toeplitz matrix analysis 
reduces to that of a scalar Toeplitz matrix.  

In this work, we extend further the above sequence of complexities by considering the R\'enyi 
entanglement entropy for quadratic spinless fermionic chains with complex finite range couplings.  
Due to the complex nature of the couplings, some new possibilities, to our knowledge not known 
in the literature, appear. We have to deal with some interesting challenges associated with some 
discontinuities in the symbol of a block Toeplitz matrix. As a resolution of these challenges 
we here propose a new formula for the determinant of such Toeplitz matrix with $2\times 2$ symbol (\ref{conjecture}):
\begin{equation*}
\log D_X(\lambda)=\frac{|X|}{2\pi}
 \int_{-\pi}^{\pi}\log\det[\lambda I-\mathcal{G}(\theta)]\mathrm{d}\theta
 +\log |X| \sum_{j=1}^J b_j
+\cdots,
\end{equation*}
where $b_j$ encapsulates the information due to the discontinuities of the symbol. 
The important fact of this formula is its logarithmic scaling with the size of the subsystem $|X|$. 
As an application of this new formula, we proceed to an exhaustive analysis 
of the entropy computation of a spin chain with Dzyaloshinski-Moriya (DM) coupling
and a Kitaev fermionic chain with long range pairing.                

As stated by Its, Mezzadri and Mo in \cite{Its2}, ``... At the core of our derivation of the entropy of
entanglement is the computation of determinants of Toeplitz matrices for a wide class of $2\times 2$ matrix symbols.''. 
The $2\times 2$ matrix character of these symbols stems from the 
non-invariance of the fermionic number. This is a distinctive feature of the XY model. 
We also consider finite range couplings; being the Kitaev chain with long range pairing an example. 
The novelty in our work lies in the complex nature of the couplings. 
This feature allows one to consider systems with broken reflection and/or charge conjugation. 
An example of a system with broken reflection symmetry is the DM spin chain.

When considering such systems with broken reflection and/or charge conjugation, 
we obtain a wider moduli of system, among those one may encounter systems with 
critical behaviour. Such systems are expected to display an entropy that scales 
logarithmically with the size of the system. 
This is indeed what we obtain when we use our proposed formula for the determinant of the block Toeplitz 
matrix in the R\'enyi entropy formula.

We organize this paper as follows. In section 2, we present the general Hamiltonian for a 
spinless fermionic chain with finite range coupling and define the Fock space of states; 
in section 3, we review the well known relation between the entropy
and the correlation matrix and in (\ref{symbol}) we derive the latter
for a general stationary state; 
in section 4, we compute the determinant of the block Toeplitz matrix, first in the smooth case where the 
Its, Mezzadri and Mo's formula (\ref{determinant}) applies, and then by extending it in (\ref{conjecture})
to the case where discontinuities are present; later we use this result to obtain
our general expression (\ref{entropy}) for the asymptotic behaviour of the entanglement entropy
in a critical theory with long range couplings and the above broken symmetries; in sections 5 and 6, 
we apply the formulae of previous section for an analysis of the entropy in the DM spin chain model
and in a Kitaev chain with long range pairing respectively;
finally, section 7 contains our conclusions.

\section{Finite range Hamiltonian}

Let us consider an $N$-site chain of size $\ell$ with $N$ spinless 
fermions described by a quadratic, translational invariant Hamiltonian 
with finite range couplings ($L<N/2$)
\begin{equation}\label{hamiltonian} 
 H=\frac12
\sum_{n=1}^N \sum_{l=-L}^L ~\left( 2 A_l a_n^\dagger a_{n+l}+B_l a_n^\dagger a_{n+l}^\dagger
-\overline B_l a_n a_{n+l} \right).
\end{equation}
Here $a_n$ and $a_n^\dagger$ represent the annihilation and creation operators at the site $n$. 
The only non-vanishing anticommutation relations are
 $$\{a_n^\dagger, a_m\}=\delta_{nm},$$
and in (\ref{hamiltonian}) we assume $a_{n+N}=a_n$.

In order that $H$ is Hermitian, the hopping couplings must satisfy  $A_{-l}=\overline{A}_{l}$. 
In addition, without loss of generality, we may take $B_l=-B_{-l}$.  
When $B_l=0$, $\forall l$, our Hamiltonian preserves the fermionic number. 
If $A_l$ is not real the reflection symmetry ($n\mapsto N-n$) is broken. 
The imaginary part of $B_l$ breaks charge conjugation ($a_n \mapsto a_{N-n}^\dagger$).  
 
We can express $H$ in terms of uncoupled fermion modes. 
First, given the translational invariance of the Hamiltonian, 
we introduce the Fourier modes
 $$b_k=\frac{1}{\sqrt{N}}\sum_{n=1}^N e^{-i\theta_k n} a_n; 
 \qquad \theta_k=\frac{2\pi k}{N}, \quad k\in{\mathbb Z}.$$
 Notice that by construction $b_k=b_{k+N}$. 
The Hamiltonian in (\ref{hamiltonian}) can now be written as
 $$H={\cal E}+\frac12 \sum_{k=0}^{N-1} (b_k^\dagger, b_{-k})
 \left(\begin{array}{cc} F_k & G_k \\ \overline G_k & -F_{-k} \end{array}\right) 
 \left(\begin{array}{c} b_k \\ b_{-k}^\dagger \end{array}\right),$$
 with
 \begin{equation}
	 \label{Fk-def-1}
   F_k=\sum_{l=-L}^L A_l e^{i\theta_k l},
 \end{equation}
 which is real, 
 \begin{equation}
	 \label{Gk-def-1}
   G_k=\sum_{l=-L}^L B_l e^{i\theta_k l},
 \end{equation}
which is an odd function of $k$, i.e. $G_{-k}=-G_k$, and
\begin{equation}
	\label{Evar-def-1}
  {\cal E}=\frac12\sum_{k=0}^{N-1}F_k.
\end{equation}

The matrix
$$R_k=\left(\begin{array}{cc} F_k & G_k \\ \overline G_k & -F_{-k} \end{array}\right)$$
is Hermitian and satisfies
$$R_{-k}=- 
\left(\begin{array}{cc} 0 & 1 \\ 1 & 0 \end{array}\right)
\overline R_{k} 
\left(\begin{array}{cc} 0 & 1 \\ 1 & 0 \end{array}\right).$$
Therefore we can introduce the dispersion relation $\Lambda_k$, with $\Lambda_{k+N}=\Lambda_k$, such that
the diagonalization of $R_k$ is given by
$$U_k R_k U^\dagger_k= \left(\begin{array}{cc} \Lambda_k & 0 \\ 0 & -\Lambda_{-k} \end{array}\right),$$
with $U_{k}$ a unitary matrix. The order ambiguity in the eigenvalues is fixed if we demand that
$\Lambda_k^S\equiv(\Lambda_k+\Lambda_{-k})/2\geq 0$. Thus the 
dispersion relation is univocally determined. 

We can now write the Hamiltonian in diagonal form by performing a Bogoliubov 
transformation to the new modes
$$\left(\begin{array}{c}d_k \\ d_{-k}^\dagger \end{array}\right)=
 U_k\left(\begin{array}{c} b_k \\ b_{-k}^\dagger \end{array}\right),$$
in terms of which $H$ reads
 $$H={\cal E}+\sum_{k=0}^{N-1} \Lambda_k
\left(d_k^\dagger d_k-\frac12\right).$$
 $\Lambda_k$ can be derived from the relations
$$
\Lambda_k^A\equiv\frac{\Lambda_k-\Lambda_{-k}}2
=\frac{F_k-F_{-k}}{2}\equiv F_k^A,
$$
and
$$\Lambda_k^S= \sqrt{(F^S_{k})^2+|G_k|^2}\geq 0,$$
with
$$F^S_{k}\equiv\frac{F_{k}+F_{-k}}{2}.$$
Finally we get
\begin{equation}
  \label{Dis-Rel-1}
  \Lambda_k=\sqrt{(F^S_{k})^2+|G_k|^2}+F^A_{k}.
\end{equation}

An eigenstate of $H$ is determined by a subset of occupied modes
$\mathbf{K}\subset \{0, \dots, N-1\}$.
If we denote by $\ket{0}$ the vacuum of the Fock space for the new modes, 
i.e.  $d_k\ket{0}=0$, the stationary state
 \begin{equation}\label{eigenstates}
  \ket{\mathbf{K}}=\prod_{k\in\mathbf{K}}d_k^\dagger\ket{0} 
 \end{equation}
has energy
 $$E_{\mathbf{K}}= {\cal E}+
\frac12\sum_{k\in\mathbf{K}}\Lambda_k - \frac12\sum_{k\not\in\mathbf{K}}\Lambda_k.$$
 In particular, the ground state will be obtained by filling the
modes with negative energy (Dirac sea)
 $$\ket{\hat{\mathbf{K}}}=\prod_{\Lambda_k<0} d_k^\dagger\ket{0},$$
 with energy 
 $$E_{\hat{\mathbf{K}}}= {\cal E}-
\frac12\sum_{k=0}^{N-1}|\Lambda_k|.$$

Note that if $|F^A_{k}|<\sqrt{(F^S_{k})^2+|G_k|^2}$ 
the dispersion relation is positive and the ground state 
of $H$ is $\ket{0}$. 
On the contrary, if there are some momenta with 
$|F^A_{k}|>\sqrt{(F^S_{k})^2+|G_k|^2}$ a Dirac sea develops,
hence the ground state has occupied modes and differs from
the Fock space vacuum. In this case, the ground state 
is not invariant under  reflection 
in momentum space $k\leftrightarrow N-k$. As we shall see in the 
next section, this fact has important consequences for the evaluation of the
entanglement entropy and its behaviour near the boundaries of the 
critical region.  

 \section{Entanglement entropy and correlation matrix}
 
The main goal of this paper is to study the R\'enyi
entanglement entropy for the ground state of a Hamiltonian 
of the kind analysed in the previous section.
For the moment, however, we keep the general discussion by considering a general stationary
state $\ket{\mathbf K}$ as defined in (\ref{eigenstates}).

Given a subsystem $X$ of contiguous sites of the fermionic chain, with complementary set $Y$ we 
have the corresponding factorization of the Hilbert space ${\cal H}={\cal H}_X\otimes{\cal H}_Y$. 
If the system is in the pure state $\ket{\mathbf K}$, the reduced density matrix for $X$ is obtained by 
taking the partial trace with respect to ${\cal H}_Y$, $\rho_X=\Tr_Y\ket{\mathbf K}\bra{\mathbf K}$. 
The R\'enyi entanglement entropy of $X$ is defined by 
\begin{equation}
  \label{Renyi-ent-1}
  S_\alpha(X)=\frac1{1-\alpha}\log \Tr(\rho_X^\alpha).
\end{equation}

In order to compute  $S_\alpha(X)$ we take advantage of the fact that the state 
$\ket{\mathbf K}$ satisfies the Wick decomposition property and, as it is well known \cite{Latorre, Peschel}, 
the reduced density matrix can be obtained from the two point correlation function 
\footnote{This construction of a density matrix out of correlation functions represents an 
instance of a general procedure known as GNS construction: one is given an algebra of observable and a 
positive linear functional acting on this algebra, and the outcome is the construction of a Hilbert space, 
where the algebra is represented, and a density matrix representing the positive linear functional. 
In \cite{Balachandran:2013cq,Balachandran:2013hga} it is discussed a unified approach to compute the entropy 
due to a partial observation, which is associated with a restriction to a subalgebra of observables. 
In particular, this framework allows the study of entropy respecting the statistics of particles.}.

For any pair of sites $n,m\in X$, we introduce the following correlation matrix, 
 $$(V_X)_{nm}=2\left< \left( \begin{array}{c} a_n \\ a_n^\dagger \end{array}\right)
 (a_m^\dagger, a_m)\right>-\delta_{nm}\; I =
 \left(\begin{array}{cc} \delta_{nm}-2\left<a_m^\dagger a_n\right> & 2\left<a_n a_m\right> \\ 
 2\left<a_n^\dagger a_m^\dagger\right> & 2\left<a_n^\dagger a_m\right>-\delta_{nm} \end{array}\right).$$
Notice that $V_X$ is a $2|X|\times 2|X|$ matrix.

Following \cite{Latorre, Peschel} 
one can show that the R\'enyi entanglement entropy reads
 \begin{equation}\label{entcorrel}
 S_\alpha(X)=\frac{1}{2(1-\alpha)}\Tr\log\left[\left(\frac{\mathbb{I}+V_X}{2}\right)^\alpha
 +\left(\frac{\mathbb I-V_X}{2}\right)^\alpha\right],
 \end{equation}
where $\mathbb I$ is the $2|X|\times 2|X|$ identity matrix.

From the numerical side, this formula supposes a drastic 
improvement of computational capabilities.
In fact, the dimension of $\rho_X$ is $2^{|X|}$, 
so the cost of calculating $S_\alpha(X)$ grows, in principle, exponentially 
with the length of the interval. However, $V_X$ has only dimension $2|X|$, allowing one to reach larger sizes of $X$.

\begin{figure}[h]
  \centering
    \resizebox{12cm}{4cm}{\includegraphics{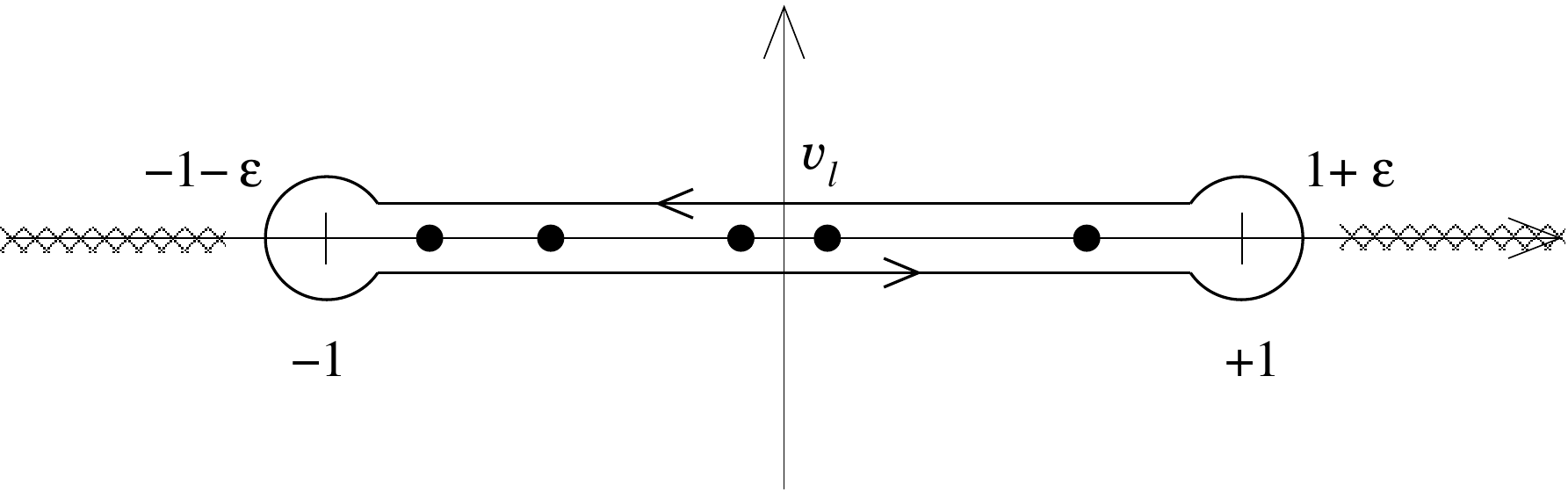}} 
    \caption{Contour of integration, cuts and poles for the computation of 
$S_\alpha(X)$. The cuts for the function $f_\alpha$ extend to $\pm\infty$.}
  \label{contorno0}
   \end{figure}

 We can derive a useful way of implementing (\ref{entcorrel}) by applying Cauchy's residue theorem, 
 so that
 \begin{equation}\label{entintegr}
 S_\alpha(X)=\lim_{\varepsilon\to0^+}\frac{1}{4\pi i}\oint_{\mathcal{C}}
 f_\alpha(1+\varepsilon, \lambda)\frac{\mathrm{d}
\log D_X(\lambda)}
{\mathrm{d}\lambda}
 \mathrm{d}\lambda,
\end{equation}
 where $D_X(\lambda)=\det(\lambda {\mathbb I}-V_X)$,
\begin{equation}\label{efealfa}
f_\alpha(x, y)=\frac{1}{1-\alpha}\log\left[\left(\frac{x+y}{2}\right)^\alpha+\left(\frac{x-y}{2}\right)^\alpha\right],
\end{equation}
 and $\mathcal{C}$ is the contour depicted in Fig. \ref{contorno0} which 
 surrounds the eigenvalues $v_l$ of $V_X$, all of them lying in the real interval $[-1,1]$. 
   
In order to compute $V_X$ we first rewrite it in the Fourier basis,
$$(V_X)_{nm}=\frac{1}{N}\sum_{k=0}^{N-1} \mathcal{G}_k e^{i\theta_k (n-m)},$$
with
$$\mathcal{G}_k=2\left< \left( \begin{array}{c} b_k \\ b_{-k}^\dagger \end{array}\right)
 (b_k^\dagger, b_{-k})\right>-I.$$ 
Or in terms of the Bogoliubov modes
$$\mathcal{G}_k=2U^\dagger_k\left< \left( \begin{array}{c} d_k \\ d_{-k}^\dagger \end{array}\right)
 (d_k^\dagger, d_{-k})\right>U_k-I. 
$$ 
We now compute the expectation value in the stationary state
$\ket{\mathbf K}$ associated to the occupation 
${\mathbf K}\subset\{0,1,\dots,N-1\}$, so that 
\begin{equation}
\left<\left( \begin{array}{c} d_k \\ d_{-k}^\dagger \end{array}\right)
 (d_k^\dagger, d_{-k})\right>=\left(\begin{array}{cc}1-\chi_{_{\mathbf K}}(k)
&0\\0&\chi_{_{\mathbf K}}(N-k)\end{array}\right),
\end{equation}
where $\chi_{_{\mathbf K}}(k)$ is the characteristic function of ${\mathbf K}$, i.e.
it is 1 or 0 according to whether $k$ belongs to ${\mathbf K}$ or not.
Now, introducing
$$M_k\equiv
U_k^\dagger\left(\begin{array}{cc}1
&0\\0&-1\end{array}\right)U_k=
\frac{1}{\sqrt{(F^S_{k})^ 2+|G_k|^2}}
 \left(\begin{array}{cc} F^S_{k} & G_k \\ \overline G_k & -F^S_{k} \end{array}\right),$$
we finally arrive at
\begin{equation}\label{symbol}
\mathcal{G}_k=\left\{\begin{array}{rc}-M_k,&\quad \mbox{if} \quad
 k\in\mathbf{K}\quad \mbox{and} \quad N-k\in\mathbf{K}, \\
 -I, &\quad \mbox{if} \quad
 k\in\mathbf{K}\quad \mbox{and} \quad N-k\not\in\mathbf{K}, \\
 M_k, &\quad \mbox{if} \quad
 k\not\in\mathbf{K}\quad \mbox{and} \quad N-k\not\in\mathbf{K},\\
 I, &\quad \mbox{if} \quad
 k\notin\mathbf{K}\quad \mbox{and} \quad N-k\in\mathbf{K}. \end{array}\right.
\end{equation}
Notice that the symbol of $V_X$ at momentum $k$ depends not only on the occupation
number at $k$ but also at $N-k$. This is a general fact that, to our knowledge, 
has been overlooked in the literature. We believe that there are two reasons for that. First, if $F^A(k)=0$ 
and we are interested in a ground state like that in \cite{Its, Its2}, then
we have ${\mathbf K}=\emptyset$. Hence any consideration on the occupation numbers is superfluous. The symbol is $M_k$ and that is all. 
The second situation corresponds to general stationary states, like in 
\cite{Ares1,Ares2} where the whole casuistics in the computation
of the symbol would play a role. However in these papers, like in 
\cite{Alba} one has $G_k=0$, therefore $M_k$ is diagonal, the symbol 
can be considered scalar instead of a 2 by 2 matrix and the discussion 
above is not necessary.

In this paper we are interested in computing $S_\alpha(X)$ for the ground state. 
As it is well known, it contains information on the critical properties of the system. 
Following the previous discussion, it is determined by the occupation set
$$\hat{\mathbf K}=\{k\,|\Lambda_k<0\}.$$
Since $\Lambda_k=\Lambda^S_k+F^A_k$, with 
$\Lambda^S_k=\Lambda^S_{N-k}\geq 0$ and $F^A_k=-F^A_{N-k}$,
applying the general expression in (\ref{symbol}), we obtain
\begin{equation}\label{symbolGS}
\hat{\mathcal G}_k=\left\{\begin{array}{rlcccc}
-I, &\quad \mbox{if} 
\quad -\Lambda^S_k&>&F_k^A,\\
 M_k, &\quad \mbox{if} 
\quad -\Lambda^S_k&<&F_k^A&<&\Lambda_k^S,\\
 I, &\quad \mbox{if} \quad ~~~F_k^A&>&\Lambda_k^S.\\
\end{array}\right.
\end{equation}
Note that for the vacuum state, and due to the fact that $\Lambda_k^S\geq 0$,
one can not have that $\Lambda_k$ and $\Lambda_{N-k}$ are both negative
implying that the first case in (\ref{symbol})  never happens.

In the so-called thermodynamic limit, $N\to\infty$, $\ell\to\infty$ with $N/\ell$ fixed, the previous 
$N$-tuples, like $\Lambda_k, F_k, G_k$ and 
the others, become $2\pi$-periodic functions determined 
by the relation $\Lambda(\theta_k)=\Lambda_k$, $\theta_k=2\pi k/N$
and analogously for $F(\theta), G(\theta),\dots$.

\section{Block Toeplitz determinant and entanglement entropy}

In this section we compute an expression for $D_X(\lambda)\equiv \det(\lambda {\mathbb I}-V_X)$ for large $|X|$. 
This enters in the formula (\ref{entintegr}) so that we can obtain the asymptotic behaviour of
$S_\alpha(X)$ for large $|X|$.

There exists a generalization for block Toeplitz matrices of Szeg{\"o} theorem \cite{Widom}
that allows to compute the linear dominant term of $D_X(\lambda)$. According to it,
 \begin{equation}\label{linear}
\log D_X(\lambda)=\frac{|X|}{2\pi}
 \int_{-\pi}^{\pi}\log\det[\lambda I-
 \hat{\mathcal{G}}(\theta)]\mathrm{d}\theta
 +\cdots,
\end{equation}
where the dots denote sub-linear contributions. If we apply this general result to our case, 
the integrand with $\hat{\mathcal{G}}(\theta)$ read from (\ref{symbolGS}) becomes the simple result:
\begin{equation}
\det[\lambda I-
 \hat{\mathcal{G}}(\theta)]=\left\{\begin{array}{rlcccc}
(\lambda+1)^2, &\quad \mbox{if} 
\quad -\Lambda^S_k&>&F_k^A,&&\\
\lambda^2-1, &\quad \mbox{if} 
\quad -\Lambda^S_k&<&F_k^A&<&\Lambda_k^S,\\
 (\lambda-1)^2, &\quad \mbox{if} 
\quad ~~~F_k^A&>&\Lambda_k^S.\\
\end{array}\right. 
\end{equation}
After inserting this result into (\ref{linear}) and then (\ref{entintegr}) we immediately see that
the linear, dominant contribution to the entropy vanishes. Therefore, in order to determine 
its asymptotic behaviour one is forced to compute the sub-dominant terms for 
the determinant that are hidden in the dots of (\ref{linear}). 

\subsection*{Case 1: $F^A(\theta)^2<F^S(\theta)^2+|G(\theta)|^2$}

We first discuss the case in which 
$F^A(\theta)^2<F^S(\theta)^2+|G(\theta)|^2$ for all $\theta\in(-\pi,\pi]$. The symbol $\hat{\cal G}(\theta)$ is therefore continuous.
 
This situation has been analysed before in the notable papers \cite{Its,Its2} when $F(\theta)=F(-\theta)$ and $G(\theta)$ is imaginary. 
Their results can be straightforwardly generalised to our 
case provided 
\begin{equation}\label{condG}
\overline{G(\theta)}=-{\rm e}^{i\psi}G(\theta).
\end{equation}

The generalisation goes as follows. We first define the meromorphic complex 
functions
$$\Phi(z)=\sum_{l=-L}^LA_l z^l,\quad \Xi(z)=\sum_{l=-L}^LB_l z^l,$$
that are related to $F$ and $G$, from (\ref{Fk-def-1}) and (\ref{Gk-def-1}), by $F(\theta)=\Phi({\rm e}^{i\theta})$
and $G(\theta)=\Xi({\rm e}^{i\theta})$.
We also introduce the symmetric and antisymmetric part of $\Phi$,  
$\Phi^S(z)=\frac12(\Phi(z)+\Phi(z^{-1}))$, 
$\Phi^A(z)=\frac12(\Phi(z)-\Phi(z^{-1}))$.  
Since $\overline A_l=A_{-l}$ and $B_{-l}=-B_l$, 
these functions satisfy
$$\Phi^S(z^{-1})=\Phi^S(z),\quad \overline{\Phi^S(z)}=\Phi^S(\overline z),
\quad \Xi(z^{-1})=-\Xi(z),$$
and this, together with (\ref{condG}), implies
$$\overline{\Xi(z)}={\rm e}^{i\psi}\Xi(\overline z).$$
We now define the polynomial
\begin{eqnarray}
P(z)=z^{2L}\left(\Phi^S(z)+{\rm e}^{i\psi/2}\Xi(z)\right)
\left(\Phi^S(z)-{\rm e}^{i\psi/2}\Xi(z)\right) 
=K\prod_{j=1}^{4L}(z-z_j),
\end{eqnarray}
with $K=\frac14(A_L+\overline{A_L})^2-|B_L|^2$.
From the symmetries of $\Phi$ and $\Xi$
one can show that $\Phi^S(z)\pm{\rm e}^{i\psi/2}\Xi(z)$ 
have real coefficients. Moreover, the roots of
$P(z)$ come in pairs $z_j$ and $z_j^{-1}$.

We include the possibility of a global phase $\psi$
for completeness, but actually it is easy to get rid of it by performing 
a redefinition of the creation and annihilation operators. 
In fact, by introducing $a'_n={\rm e}^{i\psi/4}a_n$ 
the new meromorphic functions are $\Phi'=\Phi$, $\Xi'={\rm e}^{i\psi/2}\Xi$,
therefore
$$\overline{\Xi'(z)}=\Xi'(z),$$
and the global phase is absent.

We assume that $P(z)$ is generic in the sense that it has $4L$ 
different single roots. Then the complex curve
\begin{equation}\label{Riemann}
w^2=P(z)
\end{equation}
defines a Riemann surface of genus $g=2L-1$.

As it is shown in \cite{Its2}, in the limit of large
$|X|$, $D_X(\lambda)$ can be expressed in terms of analytic invariants of the 
Riemann surface $\mathbb{C}^g$. Namely, we make a choice of branch cuts in the $z$ plane 
that do  not cross the unit circle and  a basis of fundamental cycles such
that every $a$-cycle surrounds anticlockwise one of the cuts with the last
$L$ of them corresponding to the cuts outside the unit circle.
Associated to this choice we have a canonically defined theta function,
$\vartheta:{\mathbb C}^g \rightarrow{\mathbb C}$, given by 
$$\vartheta(\vec s)\equiv \vartheta(\vec s|\Pi)=\sum_{\vec{n}}~e^{i\pi\vec n\cdot\Pi\vec n+
2i\pi\vec s\cdot\vec n}, \qquad \vec{n} \in \mathbb{Z}^g,$$
where $\Pi$ is the $g\times g$ period matrix. 
Now, one has the following expression for the asymptotic value of 
the determinant (Proposition 1 of \cite{Its2}):
\begin{equation}\label{determinant}
\log{D_X(\lambda)}
=
{|X|\log(\lambda^2-1)}+\log\frac{
\vartheta(\beta(\lambda)\vec e +\frac{\vec\tau}2)
\vartheta(\beta(\lambda)\vec e -\frac{\vec\tau}2)
}{\vartheta(\frac{\vec\tau}2)^2}+\cdots,
\end{equation}
where the dots represent terms that vanish in the large $|X|$ limit.
The argument of the theta function contains
\begin{equation}\label{beta}
\beta(\lambda)=\frac1{2\pi i}\log\frac{\lambda+1}{\lambda-1},
\end{equation}
$\vec{e}\in\mathbb{Z}^g$ which is a $g$ dimensional vector whose first $L-1$ entries are 0 and
the last $L$ are 1 and the constant $g$-tuple $\vec\tau$ is
determined once we fix which of the roots of $P(z)$ are also roots
of its first factor $z^L(\Phi^S(z)+{\rm e}^{i\psi/2}\Xi(z))$
(see Ref. \cite{Its2} for details).

\subsection*{Case 2: $F^A(\theta)^2>F^S(\theta)^2+|G(\theta)|^2$}
 
We move on now to discuss the other case, in which
$F^A(\theta)^2>F^S(\theta)^2+|G(\theta)|^2$ for some 
open intervals in $(-\pi,\pi]$. The symbol now has discontinuities at the boundaries of the interval 
where $F^A(\theta)^2=F^S(\theta)^2+|G(\theta)|^2$. 

Up to now, in the presence of such discontinuities none of the known results on block 
Toeplitz matrices can be applied to obtain the asymptotic behaviour for the determinant. 
We here propose a procedure that covers this gap in the literature, 
so that one can compute the asymptotics of the determinant 
in this case.

Before presenting the new procedure, let us recap some facts about Toeplitz 
matrices with scalar symbol with discontinuities and the corresponding Fisher-Hartwig theorem \cite{Basor}.

We therefore consider a genuine Toeplitz matrix, i.e. one with a scalar symbol $g(\theta)$. 
Furthermore, we assume $g(\theta)$ is a piecewise smooth symbol with jump discontinuities at 
$\theta_j,\ j=1,\dots,J$ and lateral limits $t_j^{-},t_j^{+}$
at the discontinuities. Then from the discontinuities we get a logarithmic contribution to $D_X(\lambda)$,
whose asymptotic expansion reads
 \begin{equation}\label{Super-Duper-formula-1}
\log D_X(\lambda)=\frac{|X|}{2\pi}
 \int_{-\pi}^{\pi}\log[\lambda -g(\theta)]\mathrm{d}\theta
 +\log|X| \sum_{j=1}^J
\frac1{4\pi^2}\left(\log\frac{\lambda-t_j^{-}}
{\lambda-t_j^{+}}\right)^2+\cdots,
\end{equation}
where now the dots represent finite contributions in the 
large $|X|$ limit, that can be computed explicitly but
are not going to be necessary for us.
The crucial fact is that contrary to the linear term
(and also the constant one) the logarithmic contribution
only depends on the behaviour of the symbol
near its discontinuities, more concretely
on its lateral limits. 

The previous observation can be used to compute
the logarithmic term for a block Toeplitz matrix in
some particular cases.
Assume that a two-dimensional symbol
$\mathcal{G}(\theta)$ presents a discontinuity
at say $\tilde\theta$.
We shall assume that the lateral limits, 
$\mathcal{G}^-=
\lim_{\theta\to\tilde{\theta}^-}\mathcal{G}(\theta)$ 
and $\mathcal{G}^+=
\lim_{\theta\to\tilde{\theta}^+}\mathcal{G}(\theta)$,
commute. If the two matrices commute there is a basis where 
both are diagonal.
Call $\tau^\pm_1$ and $\tau^\pm_2$ the corresponding
eigenvalues at each side of the discontinuity. 
Inspired by the Fisher-Hartwig theorem for the scalar case,
we argue that the contribution of this discontinuity
to the logarithmic term of the determinant only depends on the value 
of $\mathcal{G}(\theta)$ at each side of the jump or more 
concretely on its eigenvalues. Explicitly
the contribution of the discontinuity
to the coefficient of the logarithmic term of 
$\log D_X(\lambda)$ is
 \begin{equation}\label{logcoeff}
b=\frac1{4\pi^2}\left[\left(\log\frac{\lambda-\tau_1^{-}}
{\lambda-\tau_1^{+}}\right)^2+
\left(\log\frac{\lambda-\tau_2^{-}}
{\lambda-\tau_2^{+}}\right)^2\right].
\end{equation}
Now if we have several discontinuities with commuting lateral limits
there is a logarithmic contribution $b_j$, like the one above,
for each of them. Then, the asymptotic form of the block Toeplitz 
determinant up to finite contributions should be
 \begin{equation}\label{conjecture}
\log D_X(\lambda)=\frac{|X|}{2\pi}
 \int_{-\pi}^{\pi}\log\det[\lambda I-\mathcal{G}(\theta)]\mathrm{d}\theta
 +\log |X| \sum_{j=1}^J b_j
+\cdots.
\end{equation}
This result, which is crucial for the computation of the asymptotic behaviour
of the entanglement entropy for the general free fermionic model with finite range coupling,
was not previously known in the literature. It is one of the main
new contributions of this work. It would be very interesting to extend this to the case when the 
two lateral limits do not commute. So far we have been unable to accomplish this goal.
Fortunately, in order to study the fermionic chain that we consider
in this paper, the stated result is enough.

After this general discussion we proceed to study the asymptotic behaviour 
for the entanglement entropy in the ground state of the 
Hamiltonian (\ref{hamiltonian}) when the symbol in the thermodynamic limit
is not continuous.
The latter can be written
\begin{equation}\label{symbolGStherm}
\hat{\mathcal G}(\theta)=\left\{\begin{array}{rlcccc}
-I, &\quad \mbox{if} 
\quad -\Lambda^S(\theta)&>&F^A(\theta),&&\\
 M(\theta), &\quad \mbox{if} 
\quad -\Lambda^S(\theta)&<&F^A(\theta)&<&\Lambda^S(\theta),\\
 I, &\quad \mbox{if} 
\quad ~~~F^A(\theta)&>&\Lambda^S(\theta).\\
\end{array}\right. 
\end{equation}

The first thing one should notice is that if ${\mathcal G}$
has a discontinuity at $\theta$ it has also another one at $-\theta$.
Therefore, discontinuities come in pairs except those at $\theta=0$ or $\pi$.

We will now analyse the type of discontinuities and their contribution to 
the logarithmic term.

The first situation is when the two lateral limits are  
$M(\theta)$ and $\pm I$. Obviously, the two limits commute 
and we can therefore apply (\ref{logcoeff}). The eigenvalues of $M(\theta)$ are $\pm 1$
(indeed, $\det M(\theta)=-1$ and $\Tr M(\theta)=0$), which implies that its 
contribution to the $\log |X|$ term of
(\ref{conjecture}) is
$$
b_{MI}=\frac1{4\pi^2}\left(\log\frac{\lambda+1}
{\lambda-1}\right)^2.
$$

The second type of discontinuity is when the lateral limits 
are $+I$ and $-I$, in this case the two eigenvalues $\pm 1$
are different in both sides and we get a contribution
$$b_{II}=2b_{MI}.$$

Finally it is also possible that the matrix $M(\theta)$ itself
is discontinuous. This may happen when $F^S$ and $G$ vanish for some values of $\theta$
and at least one of them goes linearly to zero. In this case the two
lateral limits have opposite sign and the contribution to the logarithmic 
term is $$b_{MM}=2b_{MI}.$$

\begin{figure}[H]
  \centering
    \resizebox{16cm}{13cm}{\includegraphics{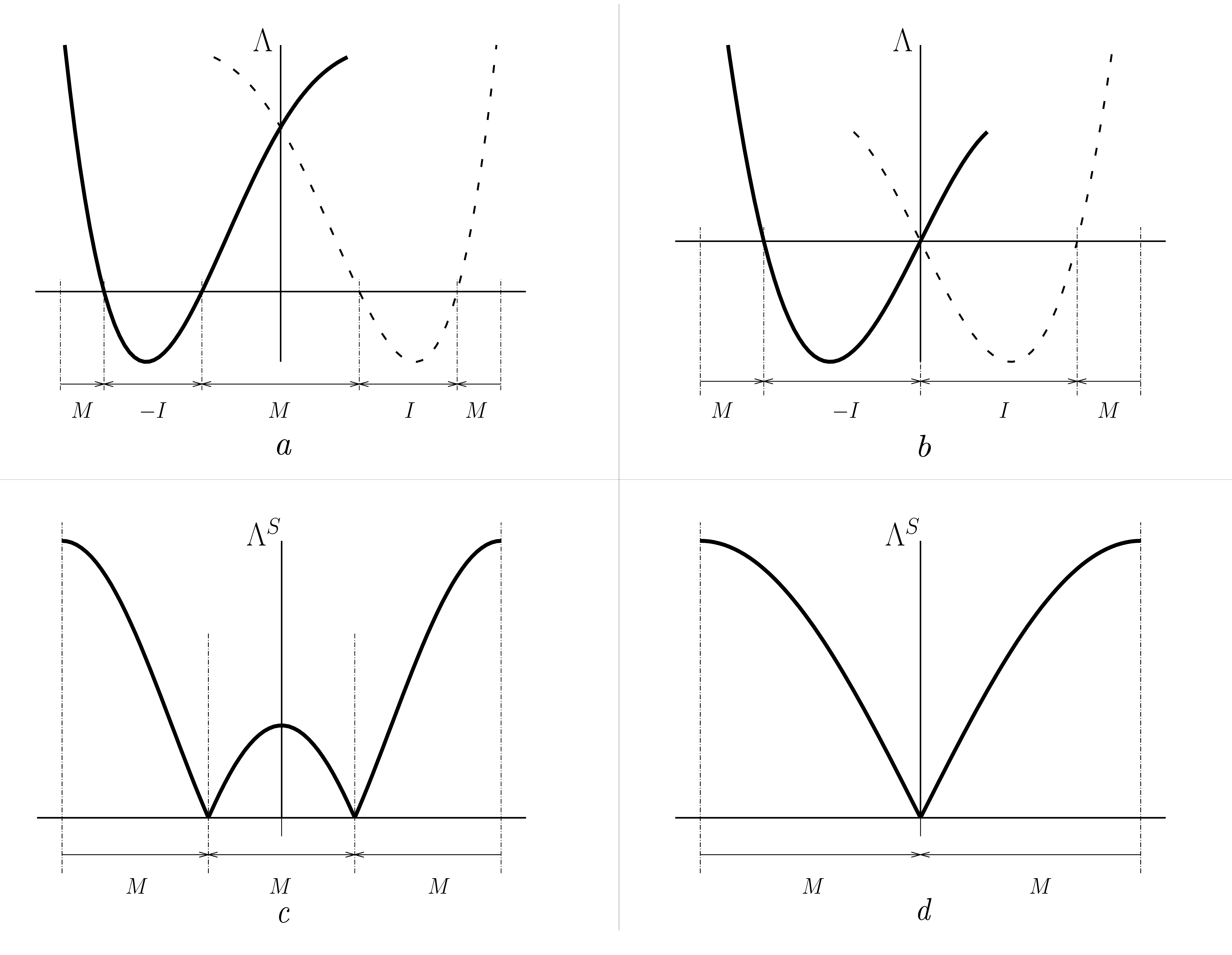}} 
    \caption{Four archetypical discontinuities for the symbol $\hat{\mathcal{G}}(\theta)$ of 
    the correlation matrix. In $a$ and $b$ the dispersion relation $\Lambda(\theta)$ is
    represented by the solid curve while the dashed curve depicts $\Lambda(-\theta)$. 
    In the plots $c$ and $d$ the solid curve stands for $\Lambda^S(\theta)$. The lines with arrows, right below the plots, 
    mark the angle $\theta$ where the discontinuities happen.}
  \label{logar_block}
   \end{figure}

From the above ingredients we can compute the asymptotic behaviour 
of the determinant for the four archetypical situations sketched
in Fig. \ref{logar_block}. 

In plot $a$ we represent a double change of sign for 
the dispersion relation for positive values of $\theta$.
We have four discontinuities of the kind $MI$,
which one can deduce from (\ref{symbolGStherm}). This is depicted with the lines 
with arrows below the plot. The contribution 
to the logarithmic term of $\log D_X(\lambda)$ is $b_a=4b_{MI}$.

In plot $b$ of Fig. \ref{logar_block}  we consider the case in which $\Lambda$ changes sign at 
$\theta=0$ (the case $\theta=\pi$ is analogous). We now have two 
discontinuities of the type $MI$ and one $II$ with a contribution
$b_b=2b_{MI}+b_{II}=4b_{MI}$. 

In the plot $c$ we represent the case in which $\Lambda^S$ vanishes
at two symmetric values of the angle, it produces two discontinuities
of the type $MM$ and hence a contribution 
$b_c=2b_{MM}=4b_{MI}$. 

Finally it may happen that $\Lambda^S$ vanishes at $\theta=0$ or $\pi$ in which case
we have only a discontinuity of the type $MM$ and a contribution
$b_d=2b_{MI}$. Altogether we have 
$$b_{T}
= 2(2n_a+2n_b+2n_c+n_d) b_{MI}\equiv N_T ~ b_{MI}.
$$
This leads to
 \begin{equation}
\log D_X(\lambda)={|X|}\log(\lambda^2-1)
 +\log |X| b_{T} 
+\cdots.
\end{equation}
Therefore, the asymptotic behaviour of the R\'enyi entanglement
entropy is 
\begin{equation}\label{entropy}
S_\alpha(X)=N_T \frac{\alpha+1}{24\alpha} \log |X| + \cdots,
\end{equation}
where we have used the identity
$$\lim_{\varepsilon\to 0^+}\frac{1}{4\pi i}
\oint_\mathcal{C} f_\alpha(1+\varepsilon,\lambda)
\frac{\mathrm{d}b_{MI}}{\mathrm{d}\lambda}\mathrm{d}\lambda=
\frac{\alpha+1}{24\alpha}.$$

 \section{Example I: Dzyaloshinski-Moriya coupling}
 
As an application of the previous results we will study the XY spin chain with
transverse magnetic field and Dzyaloshinski-Moriya (DM) coupling \cite{Kadar,Sen}.
The Hamiltonian reads
$$H_\mathrm{DM}=\frac{1}{2}\sum_{n=1}^N\left[(t+\gamma)\sigma_{n}^x\sigma_{n+1}^x+
 (t-\gamma)\sigma_n^y\sigma_{n+1}^y+s\left(\sigma_n^x 
 \sigma_{n+1}^y-\sigma_{n+1}^x\sigma_n^y\right)+h\sigma_n^z\right].$$
The coupling constants $h$, $t$, $s$ and $\gamma$ are assumed to be real. 

As it is well know, a Jordan-Wigner transform allows to write this
Hamiltonian in terms of fermionic operators. Namely
 \begin{equation}\label{dmhamiltonian}
  H_{\mathrm{DM}}=\sum_{n=1}^N 
\left[(t+is)a_n^\dagger a_{n+1}
+ (t-is)a_n^\dagger a_{n-1}
  +\gamma (a_n^\dagger a_{n+1}^\dagger-a_na_{n+1})+h a_n^\dagger a_n\right]-\frac{Nh}{2},
 \end{equation}
which can be described as a Kitaev chain with complex hopping 
couplings that break the reflection symmetry.
It is also well known that the entanglement entropy of connected
spin subchains coincide with its corresponding fermionic one and, 
therefore, for our purposes the two models are completely equivalent.

The Hamiltonian of (\ref{dmhamiltonian}) is a particular case
of (\ref{hamiltonian}) with $L=1$ and then
we can apply the last section's results.
The meromophic functions are in this case
 $$\Phi^S(z)=t z+h+t z^{-1},\quad
\Phi^A(z)=-is (z-z^{-1}),\quad
\Xi(z)=\gamma (z-z^{-1}).$$
 Therefore, the dispersion relation is
\begin{equation}\label{dispersion}
\Lambda(\theta)=\Lambda^S(\theta)
+2s\sin\theta,\quad{\rm with}\ 
\Lambda^S(\theta)=\sqrt{(h+2t\cos \theta)^2+4\gamma^2\sin^2\theta}.
\end{equation}
In the following we will fix the invariance under rescaling of 
the coupling constants by taking $t=1$. 

With a simple inspection of (\ref{dispersion}) we deduce
that the system has no mass gap when
 $$ \Delta=s^2-\gamma^2>0, \quad \hbox{and} 
 \quad \left(h/2\right)^2-\Delta<1; \quad\hbox{(Region A)},$$
 or when
 $$\Delta<0, \quad \mbox{and}\quad h=2;  \quad \hbox{(Region B)}.$$
 
\begin{figure}[h]
  \centering
    \resizebox{15.75cm}{13.125cm}{\includegraphics{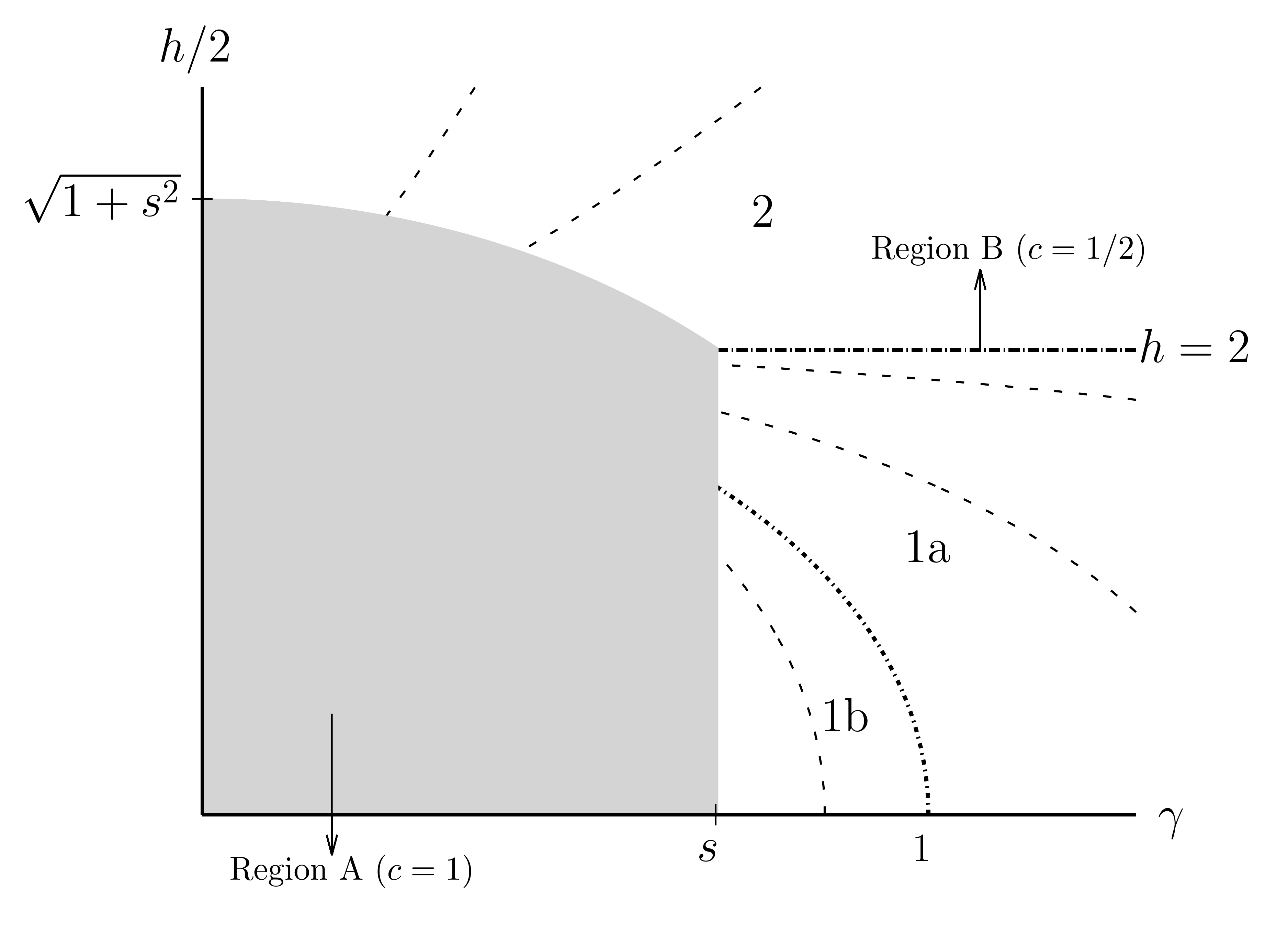}} 
    \caption{Phase diagram for $H_{\mathrm{DM}}$ in the $(\gamma, h)$ plane for $t=1$
     and fixed $s$. The shaded Region A is gapless with central charge $c=1$ (XX universality class)
     while the dashed line Region B has central charge $c=1/2$ (Ising universality class). In the unshaded
     area, the Hamiltonian has a gap and there are conical curves with the same entropy. The dashed ellipses
     and hyperbolas depict some of them.}
  \label{phase_diagram}
   \end{figure}

In Fig. \ref{phase_diagram} we depict regions A and B (actually a line)
in   the $(\gamma, h)$ plane for a fixed $s$. 

In region A the dipersion relation becomes negative in some interval
(Dirac sea) and therefore the energy is minimum when all these 
modes are occupied. It implies that the ground state corresponds
to $\hat{\mathbf K}\not=\emptyset$
and we have discontinuities in the symbol $\hat{\cal G}(\theta)$.
If $h\not=2$, it has four
discontinuities of the type $MI$, which corresponds to
the situation represented in Fig. \ref{logar_block} A. 
For $h=2$ one of the zeros of the dispersion relation is at 
$\theta=\pi$ and it corresponds to plot B in Fig. \ref{logar_block}. 
Hence, we have two discontinuities of the type $MI$ and
one of the type $II$. In both cases, applying (\ref{entropy}) 
we obtain an asymptotic behaviour for the entropy given by
$$S_\alpha(X)=\frac{\alpha+1}{6\alpha}\log|X|+\dots$$

\begin{figure}[h]
  \centering
    \resizebox{15.75cm}{10cm}{\includegraphics{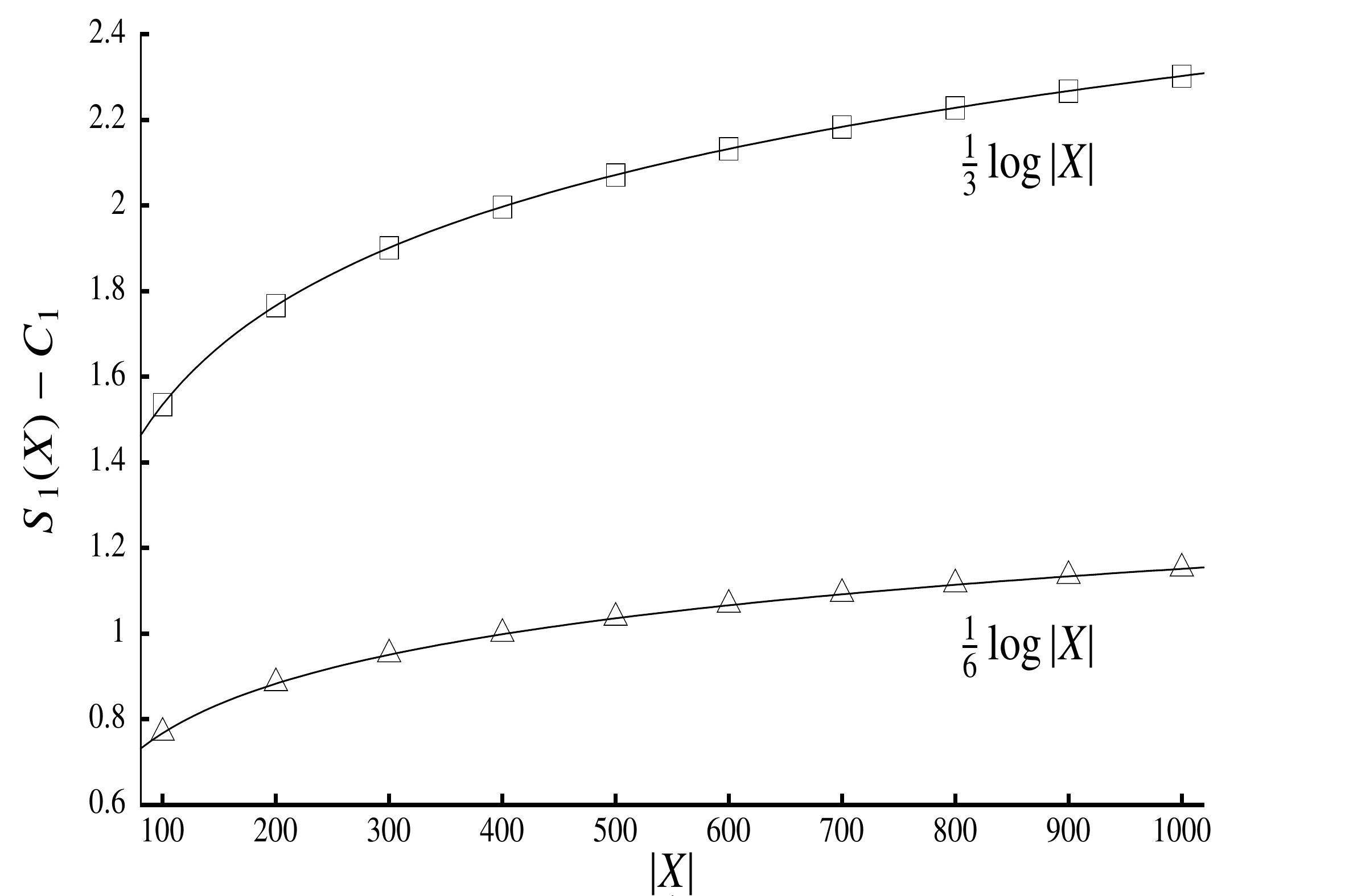}} 
    \caption{Numerical logarithmic contribution of $|X|$ to the von Neumann
     entropy ($\alpha=1$) for two sets of couplings of $H_{\mathrm{DM}}$. 
     We have subtracted the constant term $C_1=S_1(100)-c/3\log(100)$ with
     $c=1$ or $1/2$. The $\Box$ correspond to $s=0.75$, $\gamma=0.5$ and $h=0.5$, so it 
     belongs to Region A. The $\bigtriangleup$ represent $s=0.75$, $\gamma=1.5$ and $h=2$,
     so we are in Region B. The solid lines depict our
     prediction in each region.}
  \label{dm_numerics}
   \end{figure}

In the points of region B $\Lambda(\theta)$ is positive except 
at $\theta=\pi$ where it vanishes.
This implies that we have a 
single discontinuity
of the type $MM$ as we discussed before and it is represented 
in Fig. \ref{logar_block} D. 
The entropy in this region is 
$$S_\alpha(X)=\frac{\alpha+1}{12\alpha}\log|X|+\dots$$
In Fig. \ref{dm_numerics} we check numerically these scalings for the von Neumann entropy ($\alpha=1$).
   
We should discuss now the asymptotic behaviour of the entropy
for the points in the $(\gamma, h)$ plane outside the critical 
regions A and B. In this region the dispersion
relation is always positive and 
$$\hat{\cal G}(\theta)=M(\theta),$$
which besides is continuous, because $\Lambda^S(\theta)>0$. 

Notice that in our case $\gamma$ is real and therefore 
$\overline\Xi(z)=\Xi(\overline z)$, hence we can follow the 
general procedure in terms of the 
Riemann surface and theta functions that we mentioned in the 
previous section.
This has been actually carried out in \cite{Its} for the 
von Neumann entropy and in \cite{Franchini2} for the R\'enyi
entropy. The final results can be stated in a very simple way.

Let us introduce
$$x=\frac{1-\left(h/2\right)^2}{\gamma^2}.$$
Now we should distinguish three
regions outside the critical ones
1a, 1b and $2$. They are represented in Fig. \ref{phase_diagram}.
In every region the von Neumann
entropy is constant with $|X|$ and has a different expression.
\begin{itemize}
\item Region 1a: $0<x<1$, $\gamma^2>s^2$.
$$S_1=\frac16\left[\log\left(\frac{\ 1-{x\;\ }}{16\sqrt{\ x\;\ }}\right)
+\frac{\ 2(1+{x})\;\ }\pi I(\sqrt{\ 1-{x\;\ }})I({\sqrt{\ x\;\ }})\right]+\log 2.
$$

\item Region 1b: $x>1$, $\gamma^2>s^2$.
$$S_1=\frac16\left[\log\left(\frac{1-{x{^{^{-1}}}}}{16\sqrt{x{^{^{-1}}}}}\right)
+\frac{2(1+{x{^{^{-1}}}})}\pi I(\sqrt{1-{x{^{^{-1}}}}})I({\sqrt {x{^{^{-1}}}}})\right]+\log 2.
$$

\item Region 2: ${x}<0$, ${\gamma^2}(1-{x})>s^2$.
$$S_1=\frac1{12}\left[\log\left(16(2-{x}-{x}^{^{-1}})\right)
+\frac{4({x}-{x}^{^{-1}})}{\pi(2-{x}-{x}^{^{-1}})} 
I\left(\frac1{\sqrt{1-{x}}}\right)I\left(\frac1{\sqrt{1-{x}^{^{-1}}}}\right)
\right].
$$
\end{itemize}
Here $I(z)$ is the complete elliptic integral of the first kind
$$I(z)=\int_0^1 \frac{\mathrm{d} y}{\sqrt{(1-y^2)(1-z^2y^2 )}}.$$

The first observation is that the entropy only depends on the 
constant $x=(1-(h/2)^2)/\gamma^2$. This fact was first noticed in
\cite{Franchini} where the case of $s=0$ was discussed.
The curves of constant $x$ are ellipses in zones 1a and 1b 
and hyperbolas in zone 2, they are ploted in Fig. \ref{phase_diagram}.
It is clear that (if $s=0$ and region A shrinks to a line)
all curves intersect at $h=2,\ \gamma=0$
which implies that the entropy does not have a well defined 
limit at that point. It has been called the essential critical point in 
\cite{Franchini}.
From the expression for the von Neumann entropy,
one also observe a duality between the zones 1a and 1b 
by transforming $x$ into $x^{-1}$. In this way a curve in 
the region 1a is transformed into another one in 1b while
maintaining the same value for the entropy.
One can show that the curves of constant entropy 
as well as the duality also hold for R\'enyi entropy.

When we turn on the DM coupling ($s\neq 0$) the critical line at $\gamma=0$
in Fig. \ref{phase_diagram} blows up and transforms into region A. 
The curves of constant entropy in the plane $(\gamma, h)$
still persist, but only the portion of them outside region A.
On the other hand the entropy at the essential critical point
is well defined. As it is inside the critical region, it
grows like the logarithm of the size of the interval and in the limit
of large $|X|$ it becomes
$$S_\alpha(X)=
\frac{\alpha+1}{6\alpha}
\log\left(\frac{4s^2}{1+s^2}|X|\right)+{\cal I}_\alpha,
$$
where ${\cal I}_\alpha$ is a constant independent of $s$ whose expression is
$$
{\cal I}_\alpha=\frac{1}{\pi i}\int_{-1}^1 \frac{\mathrm{d} f_\alpha(1,\lambda)}{\mathrm{d}\lambda}
 \log\left[\frac{\Gamma(1/2-\beta(\lambda))}{\Gamma(1/2+\beta(\lambda)}\right]\mathrm{d}\lambda,
$$
where $\Gamma$ represents the Euler gamma function, $f_\alpha$ is defined in 
(\ref{efealfa}) and $\beta(\lambda)$ in (\ref{beta}).

Another special point is $\gamma=1$, $h=2$ where, as was noted in \cite{Kadar} by K\'adar
and Zimbor\'as, $H_{\mathrm{DM}}$ is Kramers-Wannier self-dual. In that case, the correlation
matrix reduces to a Toeplitz matrix with scalar symbol and employing the Fisher-Hartwig theorem they can obtain
not only the logarithmic contribution but also the constant term. For the von Neumann
entropy they have
$$S_1(X)=\begin{cases}\frac{1}{3}\log(2|X|)+\frac{1}{12}\log\left(1-s^{-2}\right)+\mathcal{I}_1; & s>1 \\
          \frac{1}{6}\log(4|X|)+\frac{\mathcal{I}_1}{2}; &  s\leq 1.
         \end{cases}
$$
Note that the logarithmic term agrees with our general
result.

Another interesting property of this theory is that when we approach the
critical zone B from 1a or 2, the entropy saturates at a 
larger and larger value that grows logarithmically with the correlation length 
(or the inverse of the mass gap). This is the expected behaviour that has been
predicted from the properties of conformal field theories \cite{Calabrese} and has been
verified numerical and analytically in many models.

As for the boundary of region A, it can be reached from inside the region
or from outside.
These two limits are completely different.
Of course, inside the critical region
the entropy diverges logarithmically with the 
length of the interval and, what matters for us,  
the coefficient of this term is 
constant throughout all the region.
Contrary to this behaviour, the constant coefficient in the asymptotic 
expansion of the entropy does change inside region A and it indeed 
diverges with negative values when we tend to any boundary.
If, we now study the case in which we approach zone A from any of the 
non critical ones (1a, 1b or 2) we find that the entanglement entropy 
of the interval saturates in the limit at a finite value, which is different
at any point of the boundary but independent of the path 
(always inside the non critical region) 
that we follow to reach the point.
Note that this behaviour is anomalous in the sense
that, as we mentioned before, one expects an exponential
divergence of the entropy with the inverse of the mass gap.
Recall that in the non critical regions 1a, 1b and 2 the ground
state is the Fock space vacuum for the Bogoliubov modes; the same
occurs in the critical region B. On the contrary, in region A
the ground state has all Bogoliubov modes with
negative energy occupied (the Dirac sea). The above behaviour of the
entropy near the transition could be a sign of this \textit{discontinuity}
in the ground state. In fact, we find a similar issue in the XX spin chain
(i.e. when $s, \gamma=0$). In that case, the entropy
always vanishes when $h>2$, it does not diverge with the inverse of the mass
gap when we approach the critical point $h=2$, and the ground state is the 
Fock space vacuum. On the contrary, when $h<2$, the von Neumann entropy 
scales like $1/3\log|X|$ while the constant correction diverges to negative values
when we approach the essential critical point. In this region a Dirac sea develops 
and the ground state has occupied all the modes with negative energy. The situation is,
therefore, very similar to what we obtained for the DM model.

From a global point of view,
this behaviour in the border of region A can be interpreted as a blow up of the essential critical point
of Ref. \cite{Franchini}.
The different possible values for the limit of the entropy at 
that point (for $s=0$) are obtained at different points of the boundary of 
region A for $s\not=0$ and the singularity at $h=2$, $\gamma=0$ 
disappears in this case. 
\section{Example II: Kitaev chain with long range pairing}

As a further application of the previous results we shall discuss the case in which
the couplings extend throughout the whole chain. The example is adapted from \cite{elisa}
and it serves to illustrate how to use the tools of the preceding sections to determine the 
scaling behaviour of the entropy in a theory with long range couplings.

The Hamiltonian represents a Kitaev chain with power like decaying pairing, i. e.
\begin{equation}
H_{\mathrm K}=\sum_{n=1}^N 
~\left( a_n^\dagger a_{n+1}+a_{n+1}^\dagger a_{n}+ h~a_{n}^\dagger a_{n}\right)
+\sum_{n=1}^N 
\sum_{l=-N/2}^{N/2} l|l|^{-\zeta-1} (a_{n}^\dagger a_{n+l}^\dagger - a_na_{n+l})-\frac{Nh}{2},
\end{equation}
where the exponent $\zeta>0$ characterises the dumping of the coupling
with distance. Its value will happen to be critical to determine the scaling 
behaviour of the entropy.

Following the discussion in section 4 and taking the thermodynamic limit, $N\to\infty$, we define
\begin{eqnarray}
\Phi^S(z)&=&\Phi(z)=z +h+z^{-1}\\
\Xi_\zeta(z)&=&\sum_{l=1}^\infty ({z^l-z^{-l}})~l^{-\zeta}= {\rm Li}_\zeta(z)-{\rm Li}_\zeta(z^{-1}),
\end{eqnarray}
where ${\rm Li}_\zeta$ stands for the polylogarithm of order $\zeta$.
This is a multivalued function, analytic outside the real interval $[1,\infty)$ 
and, what is going to be most important for us,  has a finite limit at $z=1$ for $\zeta>1$ while it 
diverges at that point for $\zeta<1$. 

If we introduce now
$G_\zeta(\theta)=\Xi_\zeta({\rm e}^{i\theta})$ which is a purely imaginary function, 
the dispersion relation reads
$$\Lambda(\theta)=\sqrt {(h+2\cos\theta)^2+|G_\zeta(\theta)|^2},$$
and vanishes at $\theta=\pi$ for $h=2$ and at $\theta=0$ for $h=-2$ and $\zeta>1$.
These are the two instances in which the mass gap is zero.
An interesting feature of this model is that, as we shall see, it may have logarithmic scaling 
of the entropy even when it has non zero mass gap.

Notice that  $\overline {G(\theta)}=-G(\theta)$ and therefore the formalism of 
section 4 can be applied, we also have $\Lambda^A(\theta)=0$ which corresponds to cases $c$ or $d$ in section 4.
Hence, to uncover the possible logarithmic scaling of the entropy we must simply look for
discontinuities of the matrix 
$$M(\theta)
=\frac1{\Lambda(\theta)}\left(\begin{array}{cc} h+2\cos\theta & G_\zeta({\theta})
\\ -G_\zeta({\theta}) & -h-2\cos\theta\end{array}\right).$$

One source of discontinuities comes from the zeros of the dispersion relation, this corresponds to 
the cases $c$ or $d$ discussed in section 4. In our model the only zeros appear at $\theta=0$
for $h=-2$ or $\theta=\pi$ for $h=2$ and the two lateral limits of $M$ at the discontinuity
are $\pm\sigma^y$. Therefore they contribute to the effective central charge
for the scaling of the entanglement entropy with $c=1/2$. Note that this contribution has its origin
in the absence of mass gap, i. e. it is connected to the universality class of a conformal field theory.

The other possible discontinuities are related to the divergences of $G(\theta)$ at $\theta=0$;
these happen for $\zeta< 1$ and any value of $h$. The two lateral limits are also $\pm \sigma^y$ and 
the discontinuity contributes with $c=1/2$ to the scaling of the entropy. 
If $h=2$ this contribution must be added to the one coming from the zero of the dispersion relation
at $\theta=\pi$, therefore the total central charge is $c=1$. When $h=-2$ such an addition does not happen
as the two discontinuitites are actually the same and the total central charge is $c=1/2$.

\begin{figure}[h]
  \centering
{\includegraphics{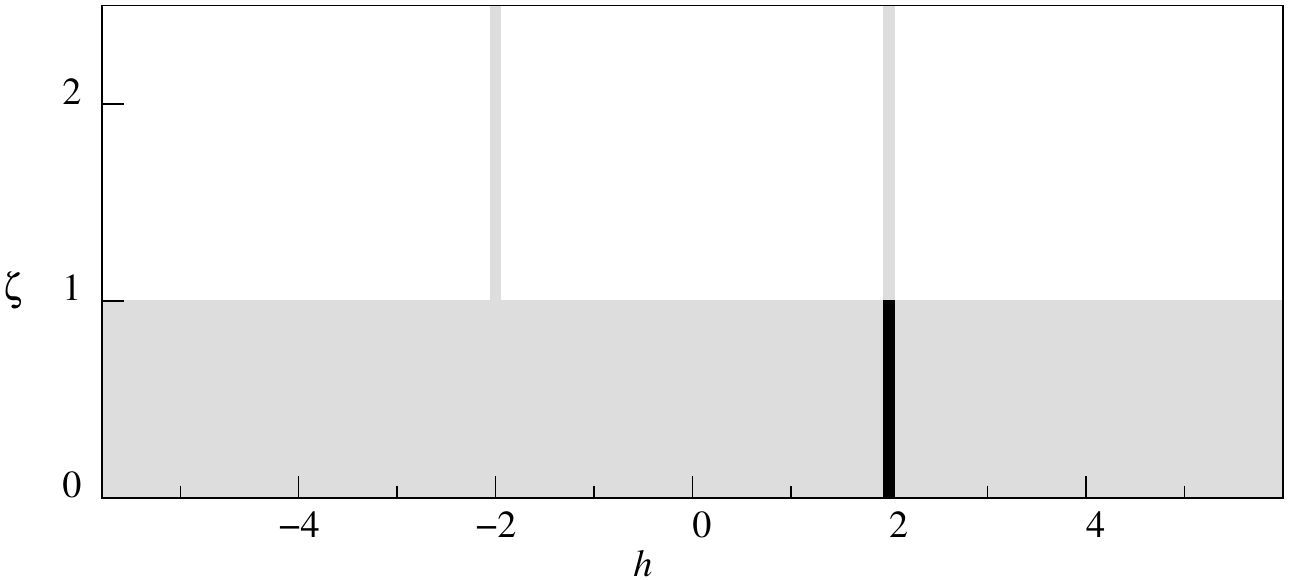}} 
    \caption{
Plot of the different regions in the $(h,\zeta)$ plane according the the 
coefficient of the logarithmic scaling for the entanglement entropy.
White area stands for $c=0$, grey region for $c=1/2$ and the black one for 
$c=1$.
}
  \label{kitaev}
   \end{figure}

All these results are summarized in Fig. \ref{kitaev} where 
the different regions, according to 
the value of $c$, are shown. The white region of central charge $c=0$ corresponds to $\zeta>1$ and $h\not=\pm2$;
in black we represent the region with $c=1$ i. e. $h=2$ and $\zeta<1$; the rest in gray corresponds to $c=1/2$.

Our analytic results are very much compatible with the numeric findings of \cite{elisa}, where the
regions we have determined appear smeared somehow. We interpret this fact as the consequence of finite size 
corrections to the  thermodynamic limit that we are presently considering.
Note that for comparison with Ref. \cite{elisa} one should bear in mind that their parameters
$\alpha$ and $\mu$ correspond to $\zeta$ and $h/2$ in our paper.

Another interesting remark is that the results above can be applied to any family of long range couplings 
with the same asymptotic behaviour for large $l$ and constant sign. 
In fact, our discussion was based on the limiting behaviour of $G$ at $\theta=0$
and the latter is governed solely by the asymptotic behaviour of the power series.

\section{Conclusions}

In this paper, we extend further the understanding of entanglement entropy
in spinless fermionic chains considering translational invariant
Hamiltonians with finite range couplings breaking 
the fermionic number, the reflection and the charge conjugation symmetries. 
The first violation implies to deal with block Toeplitz determinants while the 
second one can create Bogoliubov modes with negative energy. The latter breaking leads to discontinuities
in the symbol of our block Toeplitz matrix. To our knowledge, this possibility has not been considered in the literature. 
Here, we have performed a systematic analysis of their physical origin, nature 
and contribution to the entanglement entropy of a set of contiguous sites in the thermodynamic limit. 
For the last point, we have carried out an heuristic study based on the fact
that the contribution of each discontinuity only depends on the value of the symbol
at each side of the jump. Thereby, we conclude that each of them
adds a logarithmic term with the size of the interval to the entropy. This method also allows to determine its coefficient
which informs about the universality class of the critical theory. 
From a mathematical point of view, our result requires that the two lateral limits
of the symbol at the discontinuity commute. Although this is enough for our physical problem, it will be 
a very interesting mathematical question to study the case in which the two lateral
limits do not commute. 

When the reflection symmetry breaking does not produce Bogoliubov modes
with negative energy, the expressions found by Its, Mezzadri and Mo 
can still be applied. Here we have also shown that it can be extended to certain cases
where the charge conjugation symmetry is broken.

We have applied this general analysis to a XY spin chain with transverse magnetic field and 
Dzyaloshinski-Moriya coupling. This system has two critical regions as was previously pointed 
out in \cite{Kadar} and \cite{Sen}. Here we determine that one region belongs to the Ising 
universality class while another one region corresponds to the XX universality class. The latter can be seen as a blowing up, 
produced by the DM coupling, of the XX critical line of the XY spin chain with only magnetic coupling.

We have also tested our formula in a theory with longe range couplings, 
a Kitaev fermionic chain with power like decaying pairing. In this 
case, the logarithmic scaling of the entropy is affected by the
dumping of the coupling with the distance. For sufficiently
small values, it produces a discontinuity in the symbol of 
the correlation matrix. As a result, for zero mass gap and 
positive chemical potential, the univesality
class of the critical theory changes from the XX class (for small values of the dumping)
to the Ising class (when the dumping is sufficiently large). 
In addition, this discontinuity implies that the entropy
may have a logarithmic scaling when the mass gap is not null.
Our analytical study is in a great deal agree with the numeric results obtained
in \cite{elisa}. The difference is that
the regions we have precisely delimitated here, appear blurred in \cite{elisa}.
We interpret this fact as a consequence of finite size corrections
to the thermodynamic limit we consider here.

\noindent{\bf Acknowledgments:} Research partially supported by grants 2014-E24/2,  
DGIID-DGA and FPA2012-35453, MINECO (Spain). FA is supported by FPI Grant No. C070/2014, DGIID-DGA. 
ARQ is supported by CAPES process number BEX 8713/13-8.

 \end{document}